# Electron-Proton Decoupling in Excited State Hydrogen Atom Transfer in the Gas Phase

Mitsuhiko Miyazaki[1], Ryuhei Ohara[1], Kota Daigoku[2], Kenro Hashimoto[3], Jonathan R. Woodward[4], Claude Dedonder[5], Christophe Jouvet*[5] and Masaaki Fujii*[1]

[1] Chemical Resources Laboratory, Tokyo Institute of Technology, 4259-R1-15, Nagatsuta-cho, Midori-ku, Yokohama 226-8503, Japan

[2] Division of Chemistry, Center for Natural Sciences, College of Liberal Arts and Sciences, Kitasato University, 1-15-1 Kitazato, Sagamihara, Kanagawa 228-8555 Japan

[3] Department of Chemistry, Tokyo Metropolitan University, Minami-Osawa, Hachioji, Tokyo 192-0397, Japan

[4] Graduate School of Arts and Sciences, The University of Tokyo, 3-8-1 Komaba, Meguro, Tokyo, 153-8902, Japan.

[5] CNRS, Aix Marseille Université, Physique des Interactions Ioniques et Moléculaires (PIIM) UMR 7345, 13397 Marseille cedex, France

**Corresponding authors**

*Masaaki Fujii: mfujii@res.titech.ac.jp;*

*Christophe Jouvet: christophe.jouvet@univ-amu.fr*

**TOC**

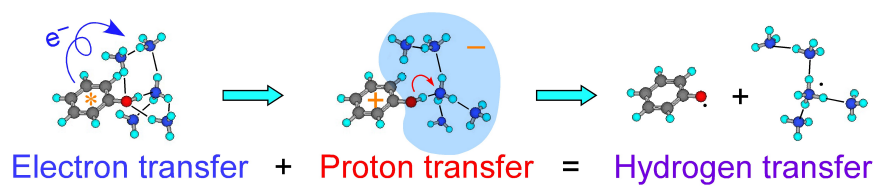

Electron transfer + Proton transfer = Hydrogen transfer

Do the hydrogen nucleus (proton) and electron move together or sequentially in the excited state hydrogen transfer (ESHT)? We answer this question by observing time-resolved spectroscopic changes that can distinguish between the electron and proton movements in a molecular cluster of phenol solvated by 5 ammonia molecules. The measurement demonstrates for the first time that the electron moves first and the proton then follows on a much slower timescale.

**Abstract**

Hydrogen-release by photoexcitation, called excited-state-hydrogen-transfer (ESHT), is one of the important photochemical processes that occur in aromatic acids. It is responsible for photoprotection of biomolecules, such as nuclei acids. Theoretically the mechanism is described by conversion of the initial state to a charge-separated state along the elongation of O(N)-H bond leading to dissociation. This means that ESHT is not a simple H-atom transfer in which a proton and a 1s electron move together. Here we demonstrate experimentally that the electron-transfer and the proton-motion is decoupled in gas-phase ESHT. We monitored electron and proton transfer processes independently by picosecond time-resolved near-infrared and infrared spectroscopy for the isolated phenol–(ammonia)$_5$ hexamer, a benchmark molecular cluster. Electron transfer from phenol to ammonia occurred in less than 3 picoseconds, while the overall H-atom transfer took 15 picoseconds. The observed electron-proton decoupling will allow for a deeper understanding and control of some important aspects of photochemistry in biomolecules.

Hydrogen and proton transfer reactions have important roles in chemistry and photochemistry [1]. In particular excited state hydrogen transfer (ESHT) in aromatic acids, proposed at the end of 20th century, has become an important new paradigm in photochemistry and related fields. For example, a significant number of photoinduced proton transfer reactions from X-H bonds have been re-defined as ESHT, including those of phenol (PhOH) [2], indole [3], tryptophan [4], aromatic amino acid cations [5] and many others [6]. Photoprotection mechanisms of biomolecules such as isolated nuclei acids of DNA are also discussed in terms of ESHT [7] as well as electron-driven proton processes [8] and conical intersections related to ring puckering [9], which make the excited state lifetime shorter. ESHT is thought to be a specific case of proton coupled electron transfer [1c, 1d].

The mechanism of ESHT has been described as a conversion from an aromatic $\pi\pi^*$ excited state to a Rydberg-like $\pi\sigma^*$ state. This conversion occurs through the crossing of the potential surfaces of the two states, and is referred to as a conical intersection due to its shape [10]. The $\pi\pi^*$ state is a strongly allowed state of the aromatic ring, and is the state initially prepared upon photoexcitation. The $\pi\pi^*$ potential energy surface crosses the repulsive surface of $\pi\sigma^*$ when the O-H or N-H bond is elongated. At the crossing point, the two potential energy surfaces are connected by a conical intersection which induces nonradiative transition from $\pi\pi^*$ to $\pi\sigma^*$. For monomers, the $\pi\sigma^*$ potential surface contains an additional crossing with that of $S_0$, and thus the photoexcited molecule is efficiently quenched by internal conversion to $S_0$ as shown in Figure 1A [10a]. Thus the short lifetime of aromatic chromophores including nucleic acid bases are interpreted as a result of ESHT [7], although the observation of $\pi\sigma^*$ is experimentally very difficult. For clusters such as PhOH–(NH$_3$)$_n$, the internal conversion to $S_0$ is mostly neglected because solvent stabilization of $\pi\sigma^*$ makes the H atom motion faster [10a]. Then the X-H bond can be simply elongated without dropping down to $S_0$, resulting in H atom transfer to the solvent moiety, e.g. PhOH–(NH$_3$)$_n$ → PhO• + •NH$_4$(NH$_3$)$_{n-1}$ (see Figure 1B). Originally photoexcited PhOH had been believed to release a proton, i.e. excited state proton transfer. However gas phase laser spectroscopy coupled with mass spectrometry enabled us to detect the reaction product directly, and provided evidence that the photoexcited PhOH–(NH$_3$)$_n$ generates a •NH$_4$(NH$_3$)$_{n-1}$ radical instead of NH$_4^+$(NH$_3$)$_{n-1}$ cation [2a, 2d, 2e]. This means that the most of photon energy is used to cleave the O-H bond homolytically and evaporation of solvent molecules is well suppressed [11]. Experimentally, H atom transfer in clusters provides a clear signature of ESHT, because the stable reaction product, for example •NH$_4$(NH$_3$)$_{n-1}$, has a surplus electron in a 3s Rydberg orbital, which 1) gives it a low ionization energy, 2) allows for 3p-3s Rydberg transitions in the near infrared (NIR) region, and 3) exhibits characteristic vibrational transitions [2c-f]. These characteristics provided the first evidence for ESHT [2a] and the NIR and IR spectroscopy of the products provided clear confirmation [2d, 2e, 2j]. Here, in both monomers and clusters, the key to this photochemistry is the contribution of the πσ* state.

The above interpretation of ESHT is far from the picture imagined from the words "hydrogen atom transfer", in which the proton dressed with a 1s electron moves as a H atom. In contrast, the σ* orbital is a Rydberg-like diffuse one, mostly centered on the solvent moiety and thus the σ* occupation has charge transfer character [10]. When ESHT is completed, the σ* electron becomes the pure 3s-Rydberg electron of the radical product. The corresponding proton moves from the -OH group to the solvent moiety independently from the electron transfer (see Figure 1C). Therefore, the electron motion and proton transfer are essentially decoupled in ESHT [12]. The electron-proton decoupling is the fundamental issue in the mechanism of ESHT, however to date it has not been revealed experimentally.

To detect the electron-proton decoupling in ESHT, we designed a time-resolved spectroscopic approach for PhOH–(NH$_3$)$_n$ clusters, which are a benchmark system for the study of ESHT [2j]. When ESHT takes place in PhOH–(NH$_3$)$_n$, 1) a new N-H bond is generated to form •NH$_4$ and 2) the electron moves to occupy the 3s-Rydberg orbital of the solvent moiety via the Rydberg-like σ* orbital. Feature 1) can be observed by IR spectroscopy in the 3 μm region, by the characteristic N-H stretching vibration of •NH$_4$(NH$_3$)$_{n-1}$. Thus the appearance of the signal due to NH stretching vibrations corresponds to the completed transfer of the proton and the electron from phenol to the ammonia moiety. Feature 2) can be monitored by the NIR absorption due to Rydberg-Rydberg transitions originating from the electron in the Rydberg-like σ* and / or 3s-Rydberg orbitals. The Rydberg-Rydberg transitions are <u>more than one-order of magnitude stronger than the valence ones</u> in general [2c, 2f, 13] and so are easily distinguished from valence transitions. For example, the oscillator strength of the 3p-3s and 3p-σ* transitions are calculated at ~0.4 and ~0.2 which is more than 200 times stronger than the valence transitions (see Supporting Information). Therefore the appearance of the NIR absorption signal corresponds to the movement of the electron to the Rydberg orbital located on the solvent moiety – i.e. to completed electron transfer. Both are essentially independent events, and therefore the comparison of the time-evolution of the NIR and IR transitions provides a means to determine whether electron and proton transfers are decoupled or not.

Each measurement requires two UV lasers and either one NIR or one IR laser (see Figure S1(A) in the Supporting Information) [2j]. The 1st UV laser excites the cluster and triggers ESHT. Then, 200 ns after the excitation, the 2nd UV laser ionizes the final reaction product •$NH_4(NH_3)_{n-1}$ which allows the determination of the population of photoexcited clusters. The IR / NIR laser is introduced at a delay of $\Delta t$ ps relative to the first UV laser and scanned over either the vibrational region or the 3p-3s Rydberg transition. When the IR / NIR laser is resonant with the transition of the transient species, the cluster is dissociated by predissociation. Thus the effect of the IR / NIR transition can be detected by the corresponding decrease in the size of the ion signal $NH_4^+(NH_3)_{n-1}$ produced by the second UV laser pulse. To obtain time-resolved spectra, we used picosecond tunable lasers for the excitation UV and IR / NIR lasers (3 ps, 12 cm$^{-1}$ resolution), while the 3rd harmonic of a nanosecond YAG laser was used for the UV ionization pulse. PhOH–$(NH_3)_n$ was generated by a supersonic jet expansion of a gaseous mixture of PhOH vapor, $NH_3$ gas, and He carrier gas [2j]. The detail of the experimental setup is described in both the Methods section and in the Supporting Information.

$S_1$-$S_0$ electronic transitions of PhOH–$(NH_3)_n$ have been reported previously [2d]. For clusters with $n \leq 4$, the spectra correspond to well-resolved structures, consistent with a slow ESHT reaction (24 ps for $n = 3$) [2j]. The spectral features are clearly different in the case of PhOH–$(NH_3)_5$ and only a broad absorption can be seen [2d]. This strongly suggests a fast ESHT reaction. For this reason this work focuses on the time-resolved measurements for the $n = 5$ cluster.

PhOH–$(NH_3)_5$ has a bicyclic hydrogen-bonded structure in the ground state $S_0$ (Figure 2A) [2e]. When the OH bond in PhOH is cleaved by ESHT, the H atom is transferred to the ammonia molecule which is directly H-bonded to the OH group. Thus ESHT must selectively generate the reaction product •$NH_4(NH_3)_4$ with a structure of $C_{3v}$+1 symmetry (see Figure 2A) [2f]. Such a geometrical restriction was also found in the ESHT of $n = 3$. After ESHT is complete, the reaction product undergoes isomerization to generate a more stable structure with $T_d$ symmetry. This final product can be measured in the NIR region using nanosecond laser excitation and is presented in Figure 2B as the black curve at the bottom. This strong absorption centered at around 6000 cm$^{-1}$ comes from the 3p-3s Rydberg transition [2d]. Given that any transitions from the photoexcited ππ* state in the unreacted cluster must be more than 2 orders of magnitude weaker than this signal, they are not resolved in these NIR measurements. This means that any observed NIR signal in this region must arise from a product of the reaction. Our calculations indicate that an NIR signal from the first-formed ESHT product with $C_{3v}$+1 symmetry should also give an NIR signal at a similar / slightly higher wavenumber (the calculated range is 6700–7700 cm$^{-1}$) with a comparable but weaker absorption than the final $T_d$ structure [2c, 2d]. Thus time resolved spectra in this region were recorded using a picosecond NIR probe laser in an attempt to resolve a spectral feature from the $C_{3v}$+1 product which would likely overlap with the spectrum of the $T_d$ product. The red traces (and blue traces which correspond to ten-times expanded spectra) in Figure 2B show the results of this measurement. While the signal due to the $T_d$ product grows in over a period of 200 ps, the signal at higher wavenumbers appears rapidly and decreases in size over the same period, although both signals are broad and clearly overlap. To achieve the optimum balance of maximizing signal intensity and minimizing spectral overlap, the time dependence of the signal at 8000 cm$^{-1}$ was recorded and due to its wavenumber and differing time dependence from the signal at 6000 cm$^{-1}$ the only reasonable assignment for this signal is the Rydberg transition of the $C_{3v}$+1 product.

The time-evolution of the NIR signal at 8000 cm$^{-1}$ is shown in Figure 2C. The signal was observed to rise rapidly and decay slowly. The time-evolution ±10 ps is shown in the inset with an expanded scale together with the response function of the experimental setup (black curve), which is mainly determined by the laser pulse duration of 3 ps. The rise of the signal is the same as the response function, thus we cannot determine the exact rise time. The ultrafast rise of the signal is consistent with the broad spectral feature of the UV absorption transition of PhOH–(NH$_3$)$_5$. The signal decays after the sharp rise and becomes almost constant after ~100 ps. From these results, we concluded that the observed ultrafast rise at 8000 cm$^{-1}$ corresponds to the 3p-3s and / or Rydberg-σ* transition due to electron transfer.

The theoretical calculations also predict the vibrational spectra for both the $C_{3v}$+1 and $T_d$ products in the NH stretching region. The details of the calculations are provided in the Supporting Information. The theoretical spectra shown at the bottom of Figure 3A are clearly different from one another. Thus time-resolved IR spectroscopy in this frequency region can reflect the time-evolution of the nascent ESHT product distinct from the NIR spectroscopy. Figure 3A shows the picosecond time-resolved IR spectra of PhOH–(NH$_3$)$_5$ at the delay times indicated beside the spectra. The strong free NH stretching bands of NH$_3$ at ~3200 cm$^{-1}$ and a broad H-bonded NH stretching band of •NH$_4$ at ~3000 cm$^{-1}$ gradually grow in intensity with increasing delay time, however no vibrational transition showing a sharp rise was found. To confirm the time-evolution of the nascent product, we fixed the IR laser frequency to the band at 3165 cm$^{-1}$ (indicated by the letter B in the spectra) and scanned the delay. The signal at this wavenumber should be sensitive to the presence of species formed after the proton transfer step (i.e. the C$_{3v}$+1 and T$_d$ product states). The observed time-evolution (shown in Figure 3B) did not show an ultrafast rise, and was fitted by a single exponential function with a 15 ps lifetime, indicating that the proton moves on this timescale. We also performed the same measurement for the two additional bands indicated by letters A and C, but no fast rise was detected in either band.

No fast time-evolution was observed for the NH stretching vibrational transition of the C$_{3v}$+1 product in the IR spectrum not only at the theoretically predicted frequency but also in all the observed vibrational bands, in sharp contrast to the ultrafast rise found in the NIR region. This means that chemical bond formation is not complete in the nascent ESHT product even when the NIR absorption due to the occupation of the Rydberg orbital is observed. This clearly demonstrates that the electron transfer is decoupled from the transfer of the proton in ESHT. Figure 1B summarizes the electron decoupled ESHT mechanism. The πσ* state is located very close in energy to S$_1$ ππ*, and the electron immediately occupies the σ* orbital within 3 ps after the ππ* excitation because of a strong ππ*/πσ* interaction. At this stage, the NIR transition at 8000 cm$^{-1}$ can be detected. On the other hand, the proton remains in its original location and therefore the new NH bond has not been formed: no corresponding IR transition can be observed. Subsequently, the OH bond dissociates and the Rydberg like σ* orbital becomes the 3s Rydberg orbital of the nascent C$_{3v}$+1 product. The proton takes ~15 ps to arrive at the ammonia moiety to form •NH$_4$ leading to the appearance of the characteristic NH stretching vibrations of •NH$_4$(NH$_3$)$_4$ radical with a slow rise.

In the •NH$_4$(NH$_3$)$_4$ radical, the isomerization from the C$_{3v}$+1 to T$_d$ products follows. This isomerization rate is faster than the formation of C$_{3v}$+1 because the time-evolution of vibrational transitions at B and C, which are predicted to be the vibrational bands of the $C_{3v}+1$ and $T_d$ products, are comparable (see Figure 3A). Therefore the isomerization dynamics appears not to be detected. Taking into account the short lifetime of $C_{3v}+1$, the NIR absorption at 8000 cm$^{-1}$ mainly reflects the population of the cluster in the $\pi\sigma^*$ state. Then the earlier part of the decay should reflect the proton transfer dynamics because it is the rate-determining step (see Figure 2A). The single exponential fitting with 15 ps time constant is also shown in Figure 2C. The observed decay was well reproduced by the fitting. It confirms that the rate of proton movement has a 15 ps lifetime. The slight deviation in the long delay is due to the fact that the signal also reflects the isomerization from the $C_{3v}+1$ product, although it is much faster than the proton transfer.

The background signal at long delay times in Figure 2C can be assigned to the weak contribution of the $T_d$ product. The IR spectrum of the reaction product from PhOH–(NH$_3$)$_5$ was measured using nanosecond lasers, and was assigned to the $T_d$ product based on theoretical calculations [2e]. It means that the $T_d$ product is the only species present after a long delay. The NIR spectrum at $\Delta t$ = 200 ps is very close to the spectrum measured by the nanosecond lasers, it is reasonable to assign the background signal to the tail of the very strong absorption of the $T_d$ species.

Electron and proton transfers during ESHT were monitored separately by picosecond time-resolved NIR / IR spectroscopy for PhOH–(NH$_3$)$_5$. From the different time-evolution of NIR and IR transitions, it can be concluded that the electron and proton transfers are decoupled. This is the first experimental evidence that ESHT is not a H atom transfer in the gas phase. Such a decoupling was not found in smaller clusters [2j]. The essential description of ESHT, including the contribution of the $\sigma^*$ orbital must also be the same for the smaller clusters. Tentatively, we think that proton tunneling between $\pi\pi^*$ and $\pi\sigma^*$ states synchronizes the proton motion and the electron dynamics in smaller clusters. To present a unified view of the ESHT mechanism, advanced theoretical approaches are now in progress as well as measurements on deuterated systems. This fundamental mechanism of ESHT will provide better understanding of key photochemical processes, including photoprotection of biomolecules, and will aid the design of photo-triggered molecular tools.

**Acknowledgments**
This study was supported in part by MEXT (innovative area 2503 and the Cooperative

Research Program of the "Network Joint Research Center for Materials and Devices") and JSPS (Grant-in-Aid for Scientific Research (A) 15H02157 and the Core-to-Core Program 22003). Authors acknowledge stimulating discussion with Dr. S. Ishiuchi and Prof. M. Sakai in Tokyo Institute of Technology.

**Figures and Figure Legends**

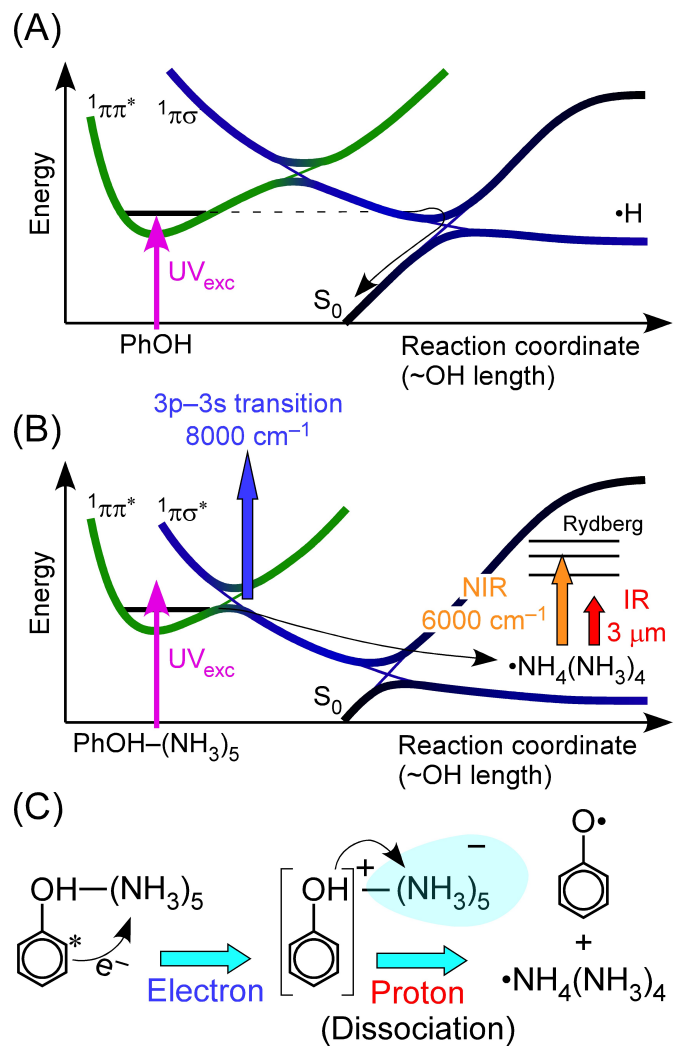

**Figure 1** (A) Schematic potential curves responsible for the ESHT reaction of phenol (PhOH) monomer and (B) for PhOH–(NH$_3$)$_5$. (C) Reaction scheme of the excited state hydrogen transfer (ESHT) reaction of phenol–(NH$_3$)$_5$.

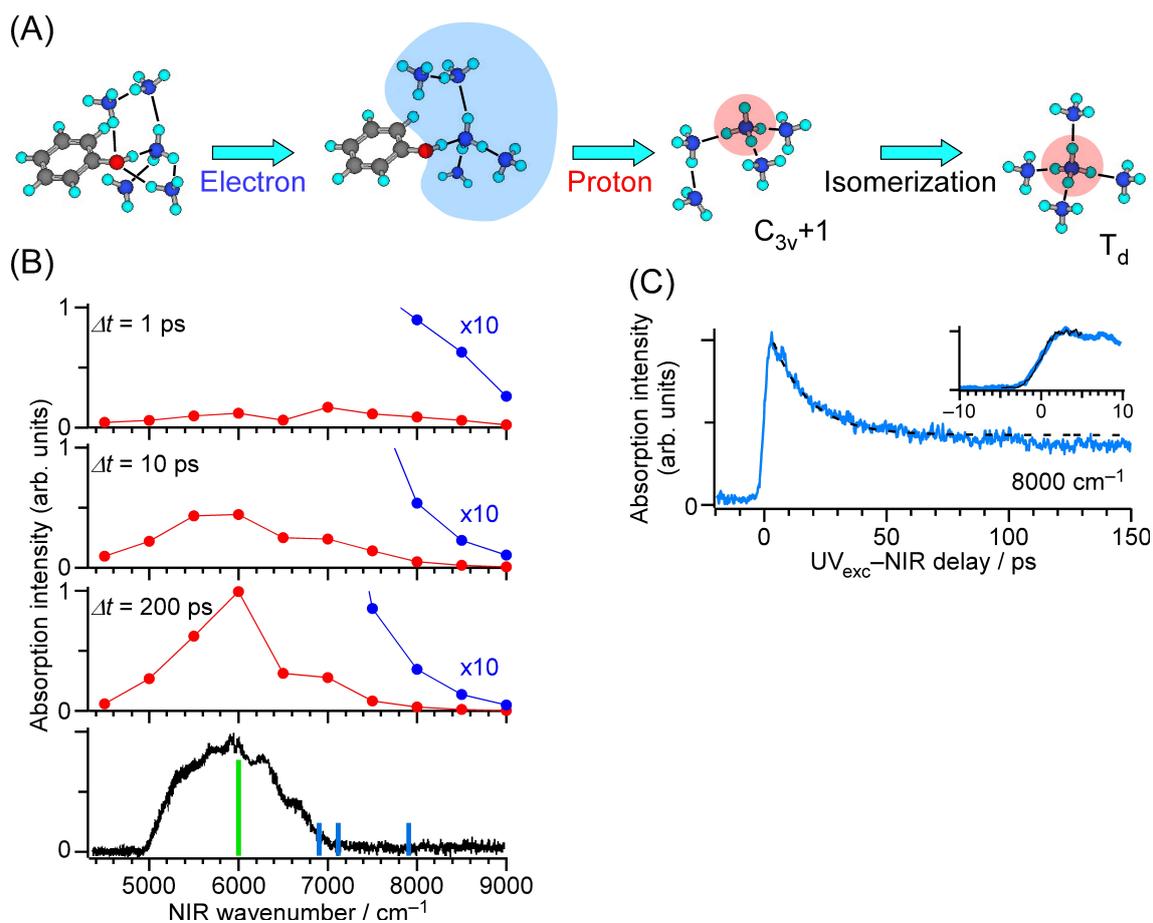

**Figure 2** (A) Structure of the PhOH–(NH$_3$)$_5$ cluster and •NH$_4$(NH$_3$)$_4$ product along the course of the ESHT reaction. (B) Picosecond time-resolved NIR spectra of PhOH-(NH$_3$)$_5$ with $\Delta t$ = 1, 10 and 200 ps (top 3 traces). The bottom trance is the NIR spectrum measured by the nanosecond lasers with $\Delta t$ = 180 ns. Green and blue bars correspond to theoretical spectra of the T$_d$ and C$_{3v}$+1 isomers, respectively. (C) Time evolution of the transient NIR absorption of PhOH–(NH$_3$)$_5$ after the ππ* excitation probed at 8000 cm$^{-1}$. The rise of the signal is expanded in the inset together with the response function of the experimental setup (black curve). The exponential fitting of a 15 ps time constant is represented by a broken line.

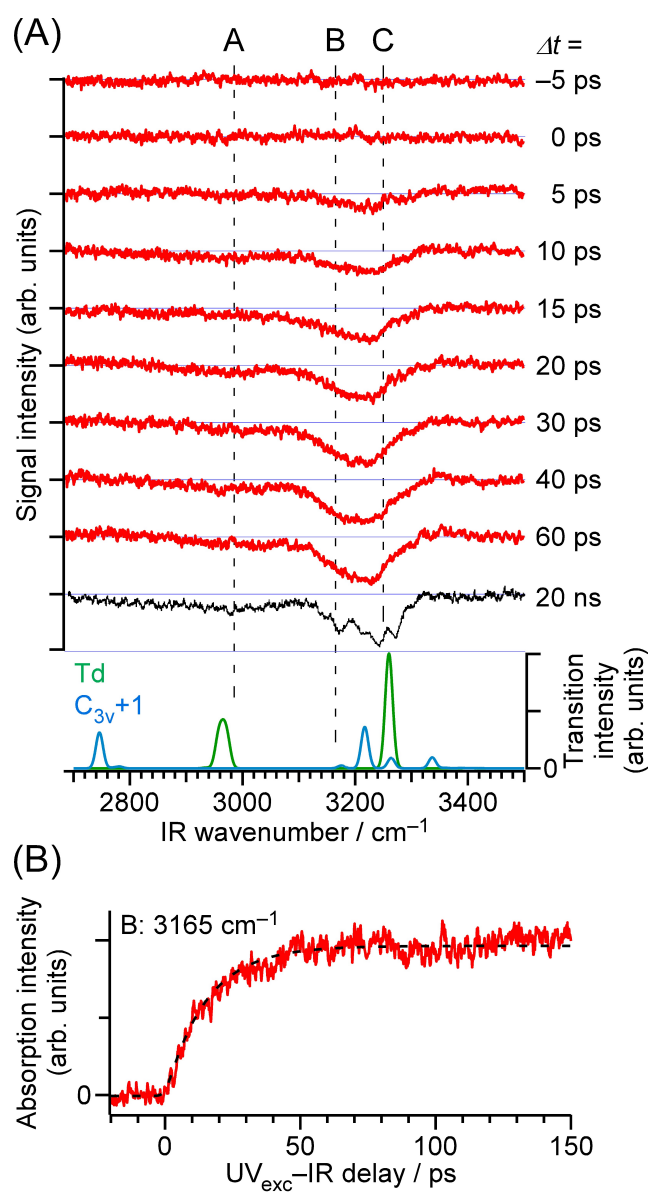

**Figure 3** (A) Time resolved IR spectra of PhOH–(NH$_3$)$_5$ after the ππ* excitation, monitored in the NH stretching vibration region. The spectrum at $\Delta t$ = 20 ns was measured using nanosecond lasers, and was adapted from the previous data published in reference 5. Theoretical spectra of T$_d$ and C$_{3v}$+1 isomers of the ESHT product, •NH$_4$(NH$_3$)$_4$, are also shown at the bottom as green and blue traces, respectively. (B) Time evolution of the NH stretching vibration probed at 3165 cm$^{-1}$ (dashed line B in figure 3 (A)). A single exponential fitting is presented as a broken curve. From the fitting, the risetime is 15 ± 2 ps.

**Supporting Information**

**(A) Evaluation of the oscillator strength of the probe photon in the mid infrared from a ππ\* or a πσ\*.**

**Table S1** Oscillator strength from the first 1a' valence state (ππ\*)

**Table S2** Oscillator strength from the first a" Rydberg state (πσ\*).

**Table S3** Cartesian coordinates of the cluster.

**(B) NH stretching**

**Table S4**

**(C) Experimental scheme and setup**

    **Figure S1**

15